\documentclass{article}

\input{tcilatex}

\begin{document}

{\footnotesize {Mathematical Problems of Computer Science {\small 23, 2004,
127--129.}}}

\bigskip

\bigskip

\begin{center}
{\Large \textbf{On Lower Bound for $W(K_{2n})$}}

{\normalsize Rafael R. Kamalian and Petros A. Petrosyan}

{\small Institute for Informatics and Automation Problems of NAS of RA}

{\small e-mails rrkamalian@yahoo.com, pet\_petros@yahoo.com}

\bigskip

\textbf{Abstract}
\end{center}

The lower bound $W(K_{2n})\geq 3n-2$\ is proved for the greatest possible
number of colors in an interval edge coloring of the complete graph $K_{2n}$.

\bigskip 

Let $G=(V(G),E(G))$ be an undirected graph without loops and multiple edges
[1], $V(G)$ and $E(G)$ be the sets of vertices and edges of $G$,
respectively. The degree of a vertex $x\in V(G)$ is denoted by $d_{G}(x)$,
the maximum degree of a vertex of $G$-by $\Delta (G)$, and the chromatic
index of $G$-by $\chi ^{\prime }(G)$. A graph is regular, if all its
vertices have the one degree. If $\alpha $ is a proper edge coloring of the
graph $G$, then $\alpha (e)$ denotes the color of an edge $e\in E(G)$ in the
coloring $\alpha $.

Let $\alpha $ be a proper coloring of edges of $G$ with colors $1,2,\ldots
,t $. $\alpha $ is interval [2], if for each color $i,1\leq i\leq t,$ there
exists at least one edge $e_{i}\in E(G)$ with $\alpha (e_{i})=i$ and the
edges incident with each vertex $x\in V(G)$ are colored by $d_{G}(x)$
consecutive colors.

A graph $G$ is interval-colorable if there is $t\geq 1$ for which $G$ has an
interval edge coloring $\alpha $ with colors $1,2,\ldots ,t$. The set of all
interval-colorable graphs is denoted by $\mathcal{N}$ [2].

For $G\in \mathcal{N}$ we denote by $W(G)$ the greatest value of $\ t$, for
which $G$ has an interval edge coloring $\alpha $ with colors $1,2,\ldots ,t$%
.

It was proved [3] for bipartite graphs that verification whether $G\in 
\mathcal{N}$ \ is $NP$-complete [4,5].

It was proved [2] that if $G$ has no triangle and $G\in \mathcal{N}$ then $%
W(G)\leq \left\vert V(G)\right\vert -1$. It follows from here that if $G$ is
bipartite and $G\in \mathcal{N}$ then $W(G)\leq \left\vert V(G)\right\vert
-1 $ [2].

For graphs which can contain a triangle the following results hold:

\textbf{Theorem 1 }[6]. If $G\in \mathcal{N}$ is a graph with nonempty set
of edges then $W(G)\leq 2\left\vert V(G)\right\vert -3$.

\textbf{Theorem 2 }[7]. If $G\in \mathcal{N}$ and $\left\vert
V(G)\right\vert \geq 3$ then $W(G)\leq 2\left\vert V(G)\right\vert -4$.

Non-defined conceptions and notations can be found in [1,6,8,9].

In [2] there was proved the following

\textbf{Theorem 3}. If $G\in \mathcal{N}$ then $\chi ^{\prime }(G)=\Delta
(G) $.

\textbf{Corollary 1 }[2]. For a regular graph $G$ \ $G\in \mathcal{N}$ \ iff
\ $\chi ^{\prime }(G)=\Delta (G)$.

A well-known inequality $\Delta (G)\leq \chi ^{\prime }(G)\leq \Delta (G)+1$
for the chromatic index of an arbitrary graph $G$ was proved in [10]. It was
proved [11] for regular graphs that verification whether $\chi ^{\prime
}(G)=\Delta (G)$ is $NP$-complete.Therefore we can conclude from the \textbf{%
Corollary 1} that the problem

\textquotedblleft Whether a given regular graph belongs to the set $\mathcal{%
N}$ or not?\textquotedblright\ is also $NP$-complete.

From the results of [10] and \textbf{Corollary 1} it follows that for any
odd $p$ $\ \ K_{p}\notin \mathcal{N}$. It is not difficult to see that $\chi
^{\prime }(K_{2n})=\Delta (K_{2n})$ (it can be obtained from the theorem 9.1
of [1]). Consequently from the \textbf{Corollary 1} we obtain

\textbf{Theorem 4}. For any $\ n\in N$ $\ \ \ K_{2n}\in \mathcal{N}$.

It is not difficult, using the results of [6], to obtain the following

\textbf{Theorem 5}. For any $n\in N$ \ \ $W(K_{2n})\geq 2n-1+\left\lfloor
\log _{2}\left( 2n-1\right) \right\rfloor $.

The main result of this paper consists of the following

\textbf{Theorem 6}. For any \ $n\in N$ \ \ $W(K_{2n})\geq 3n-2$.

\textbf{Proof}. Obviously, for $n\leq 3$ the estimate of the \textbf{Theorem
6} coincides with that one of the \textbf{Theorem 5}.

Now assume that $n\geq 4$.

Consider a graph $K_{2n}$ with $V(K_{2n})=\left\{
u_{1},u_{2},...,u_{2n}\right\} $\ and$\ $

$E(K_{2n})=\left\{ \left( u_{i},u_{j}\right) |\text{ \ }u_{i}\in V(K_{2n}),%
\text{ }u_{j}\in V(K_{2n}),\text{ }i<j\right\} $.

Define a coloring $\ \alpha $ of the edges of the graph $K_{2n}$ in the
following way:

for $\ i=1,...,\left\lfloor \frac{{\large n}}{{\large 2}}\right\rfloor ,$ $%
j=2,...,n,$ $i<j,$\ $i+j\leq n+1$ \ \ \ \ \ 

\begin{center}
$\alpha \left( \left( u_{i},u_{j}\right) \right) =i+j-2;$\ \ \ \ \ \ 
\end{center}

for \ $i=2,...,n-1,$ $j=\left\lfloor \frac{{\large n}}{{\large 2}}%
\right\rfloor +2,...,n,$ $i<j,$\ $i+j\geq n+2$ \ \ \ \ 

\begin{center}
$\alpha \left( \left( u_{i},u_{j}\right) \right) =i+j+n-3;$\ \ \ \ \ \ 
\end{center}

for \ $i=3,...,n,$ $j=n+1,...,2n-2,$ $j-i\leq n-2$\ \ \ \ \ 

\begin{center}
$\alpha \left( \left( u_{i},u_{j}\right) \right) =n+j-i;$\ 
\end{center}

for \ $i=1,...,n,$ $j=n+1,...,2n,$ $j-i\geq n$\ \ 

\begin{center}
$\alpha \left( \left( u_{i},u_{j}\right) \right) =j-i;$
\end{center}

for \ $i=2,...,1+\left\lfloor \frac{{\large n-1}}{{\large 2}}\right\rfloor ,$
$j=n+1,...,n$ $+\left\lfloor \frac{{\large n-1}}{{\large 2}}\right\rfloor ,$ 
$\ j-i=n-1$ \ \ 

\begin{center}
$\alpha \left( \left( u_{i},u_{j}\right) \right) =2(i-1);$\ \ 
\end{center}

for \ $i=$ $\left\lfloor \frac{{\large n-1}}{{\large 2}}\right\rfloor
+2,...,n,$ $\ j=n+1+\left\lfloor \frac{{\large n-1}}{{\large 2}}%
\right\rfloor ,...,2n-1,$ $\ j-i=n-1$ \ 

\begin{center}
$\alpha \left( \left( u_{i},u_{j}\right) \right) =i+j-2;$\ 
\end{center}

for \ $i=n+1,...,n+\left\lfloor \frac{{\large n}}{{\large 2}}\right\rfloor
-1,$ $j=n+2,...,2n-2,$ $i<j,$\ $i+j\leq 3n-1$ \ \ \ 

\begin{center}
$\alpha \left( \left( u_{i},u_{j}\right) \right) =i+j-2n;$
\end{center}

for \ $i=n+1,...,2n-1,$ $j=n+\left\lfloor \frac{{\large n}}{{\large 2}}%
\right\rfloor +1,...,2n,$ $i<j,$\ $i+j\geq 3n$\ 

\begin{center}
$\alpha \left( \left( u_{i},u_{j}\right) \right) =i+j-n-1.$
\end{center}

It is not difficult to see that $\ \alpha $ \ is an interval edge coloring
of the graph $K_{2n}$ with colors $1,2,...,3n-2.$

The proof is complete.

\vspace{1cm}


\bigskip \bigskip 

\begin{thebibliography}{99}
\bibitem{[1]} F. Harary, Graph Theory, Addison-Wesley, Reading, MA,1969.

\bibitem{[2]} A.S. Asratian, R.R. Kamalian, Interval colorings of edges of a
multigraph, \textit{Appl. Math}.5 , 25-34,1987.

\bibitem{[3]} S.V. Sevastianov, On interval colourability of edges of a
bipartite graph, \textit{Meth. Of Discr. Anal}. In solution of external
problems. The Institute of Mathematics of the Siberian Branch of the Academy
of Sciences of USSR. Novosibirsk, N50, 61-72, 1990.

\bibitem{[4]} S. Cook, The complexity of theorem-proving procedures. In 
\textit{Proc.3rd ACM Symp}. on Theory of Computing, 151-158, 1971.

\bibitem{[5]} R.M. Karp, Reducibility among Combinatorial Problems, in
\textquotedblleft Complexity of Computer Computations\textquotedblright\
(R.E. Miller and J.W. Thatcher, Eds.), pp. 85-103, New York, Plenum, 1972.

\bibitem{[6]} R.R. Kamalian, Interval Edge Colorings of Graphs, Doctoral
dissertation, Novosibirsk, 1990.

\bibitem{[7]} K. Giaro, M. Kubale, M. Malafiejski, Consecutive colorings of
the edges of general graphs, \textit{Discr. Math}. 236 , 131-143,2001.

\bibitem{[8]} A.A. Zykov, Theory of finite graphs, Novosibirsk, Nauka, 1969.

\bibitem{[9]} A.S. Asratian, R.R. Kamalian, Investigation on interval edge
colorings of graphs, \textit{J. Combin. Theory Ser}. B 62, 34-43,1994.

\bibitem{[10]} V.G. Vizing, The chromatic index of a multigraph, \textit{%
Kibernetika} 3, 29-39, 1965.

\bibitem{[11]} I. Holyer, The $NP$-completeness of edge coloring, \textit{%
SIAM J. Comput.} 10, N4, 718-720, 1981.
\end{thebibliography}
\end{document}